# Reduced Coulomb interaction in organic solar cells by the introduction of inorganic high-*k* nanostructured materials

**Miriam Engel[1], David Schaefer[2], Daniel Erni[2], Niels Benson[\*,1] and Roland Schmechel[1]**

[1] Institute for Nanostructures and Technology (NST), Faculty of Engineering, University of Duisburg-Essen and CENIDE – Center for Nanointegration Duisburg-Essen, D-47057 Duisburg, Germany

[2] General and Theoretical Electrical Engineering (ATE), Faculty of Engineering, University of Duisburg-Essen and CENIDE – Center for Nanointegration Duisburg-Essen, D-47057 Duisburg, Germany

\* Corresponding author: Niels Benson, e-mail: niels.benson@uni-due.de, Phone: +49 (0)203 379 1058, Fax: +49 (0)203 379 3268

In this article a concept is introduced, which allows for reduced Coulomb interaction in organic solar cells and as such for enhanced power conversion efficiencies. The concept is based on the introduction of electrically insulating, nanostructured high-*k* materials into the organic matrix, which do not contribute to the charge transport, however, effectively enhance the permittivity of the organic active layer and thereby reduce the Coulomb interaction. Using an analytical model it is demonstrated that even at a distance of 20 nm to the organic / inorganic interface of the nanostructure, the Coulomb interaction can be reduced by more than 15 %. The concept is implemented using P3HT:PCBM solar cells with integrated high-*k* nanoparticles (strontium titanate). It could be demonstrated that in comparison to a reference cell without integrated nanoparticles, the power conversion efficiencies could be improved by ~20 %.

**1 Introduction** During the last decade an increase in the power conversion efficiency of organic photovoltaic cells up to almost 11 % could be realized [1, 2]. This is mainly the result of developments in the device and material design, as well as optimized morphologies. Widely used is the bulk-heterojunction concept [3, 4] as well as tandem solar cell designs [5, 6]. However, one aspect which has received little attention during this development is the high Coulomb interaction in organic semiconductors, as a result of the low semiconductor permittivity ($\varepsilon_r \sim$ 3-4) [7, 8]. This material property contributes to high exciton binding energies of up to 1.4 eV [9], reduces charge transport properties due to electrostatic interaction and leads to enhanced geminate pair formation [3, 10, 11, 12]. An enhancement of the material permittivity would reduce the influence of the Coulomb interaction and therefore have a beneficial effect on the maximum obtainable power conversion efficiency for organic solar cells. This can be achieved either by designing novel organic semiconductor materials with an enhanced permittivity [13], or by enhancing the effective permittivity of the organic active layer. The later has been substantiated in a recent publication, by which we demonstrated facilitated exciton separation in pentacene on high-*k* substrates. This was ascribed to an effectively enhanced permittivity in the vicinity of the substrate / semiconductor interface and as such reduced electrostatic interaction between complementary charges [14]. In the current paper we demonstrate the feasibility of a concept to enhance the effective permittivity of organic solar cells, by the integration of electrically insulating, nanostructured high-*k* materials. The benefit of the organic / high-*k* interface for charge pair separation is discussed using the charge carrier escape energy. Further, first P3HT:PCBM solar cells with integrated strontium titanate ($SrTiO_3$) nanoparticles are realized, demonstrating a power conversion efficiency enhancement of ~ 17 % in comparison to a reference device.

**2 Experimental** The solar cells were realized on commercially available OLED grade glass substrates covered with indium tin oxide (ITO). Poly(3,4-ethylenedioxythiophene) poly(styrenesulfonate) (PEDOT:PSS, Clevios P VPAl 4083) was deposited via spin coating and subsequently annealed for 10 min at 165 °C. All further processing was conducted in inert $N_2$ atmosphere. The active layer for both high-*k* and reference solar cells was prepared using a 1:0.78 blend of regioregular poly(3-hexylthiophene-2,5-diyl) (P3HT, Sepiolid P200, BASF) and [6,6]-phenyl-C61-butyric acid methyl ester (PCBM, Sigma Aldrich) solved in chlorobenzene (10 mg ml$^{-1}$ P3HT). In order to be able to realize high-*k* organic solar cells, dried (150 °C under vacuum, 2 h) strontium titanate ($SrTiO_3$, Iolitec) nanoparticles were added to a part of the reference solution, which has been stirred at 65-70 °C for 8 h in a concentration of 5 mg ml$^{-1}$. Further, Ceramic grinding balls were added to the high-*k* solution for better dispersability of the $SrTiO_3$ nanoparticles. Both low-*k* solution and high-*k* dispersion were continuously stirred over night. Prior to the low-*k* solution / high-*k* dispersion deposition using a spin coater, both solution and dispersion were filtered. The low-*k* solution using a 0.7 μm filter, and the high-*k* dispersion using 2.7 μm, 1.2 μm and 0.7 filters. Metal electrodes (50 nm calcium, 150 nm aluminum) were deposited by evaporation in high vacuum. In a final step the solar cells were annealed for 10 min at 140 °C. The active area of the solar cells is 20 mm².

The devices were characterized using a Keithley source-measure unit in the dark and under standard AM1.5 illumination. The current-voltage (I-V) characteristics were measured between -1 V and +1 V under nitrogen atmosphere. Transmission measurements in a spectral range from 800 nm down to 300 nm were conducted, using a Shimadzu UV-2550 UV-VIS spectrometer in combination with an integrating sphere. Here a glass/ITO/PEDOT:PSS layer stack was used as a reference. Further, the structure of the active layer was analyzed using a JEOL scanning electron microscope.

**3 Theoretical Considerations** In order to quantify the influence of the organic / high-$k$ material interface on the electrostatic charge carrier interaction in organic semiconductors, the following analytical model is used to describe the Coulomb interaction between complementary charges at the interface between low-$k$ and high-$k$ infinite large half-spaces. The model considers a stationary charge $Q_s$ in a low-$k$ material at a distance of 2.5 nm to a low-$k$ / high-$k$ interface, as well as a reference low-$k$ / low-$k$ interface. The model investigates the interaction of a test charge with the stationary charge. Further, image charges induced at the interface by the stationary and the test charge are considered. Both test and stationary charges carry a charge equal to $|Q| = nq$.

The electrostatic interaction between the respective charges is described by the Coulomb interaction, with a force $F_{Coulomb}$ of:

$$F_{Coulomb} = \frac{Q_s Q_t}{4\pi\varepsilon_0 \varepsilon_r r^2} \tag{1}$$

Here $r$ represents the distance between the charges, $\varepsilon_r$ the homogeneous, linear, material permittivity and $\varepsilon_0$ the vacuum permittivity.

The bound charge density induced by the charges $Q_s$ or $Q_t$, is described each as a single image charge $Q'$ as discussed by Jackson [15]. The equation used to approximate the image charge $Q'$ is given in the following:

$$Q' = \frac{\varepsilon_{r,low-k} - \varepsilon_{r,high-k}}{\varepsilon_{r,low-k} + \varepsilon_{r,high-k}} Q \tag{2}$$

Equation 2 is valid for an interaction with the test charge in the low-$k$ half space. The interaction scenarios considered here are illustrated in Fig. 1. Case A (Fig. 1a) represents the reference system with a low-$k$ / low-$k$ interface. Both half spaces are assigned a permittivity of 3. For case B (Fig. 1b) a low-$k$ / high-$k$ interface is considered. Here respective permittivities of 3 and 300 are taken into account for the low-$k$ / high-$k$ material. In consequence only the Coulomb interaction between the stationary charge $Q_s$ and the test charge $Q_t$ needs to be considered for case A, while for case B also the interaction with the image charges $Q_s'$ and $Q_t'$ needs to be considered.

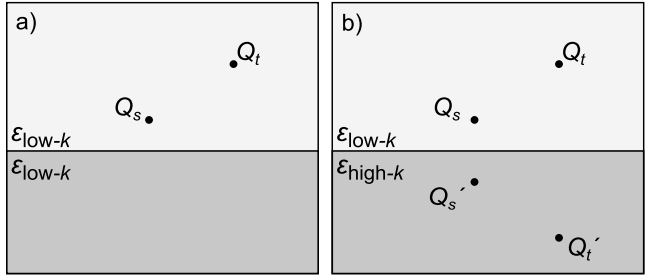

**Figure 1** (a) $Q_s$ and $Q_t$ above a low-$k$ / low-$k$ interface with a homogeneous permittivity of 3, (b) $Q_s$ and $Q_t$ above a low-$k$ / high-k interface ($\varepsilon_r$ = 3 / 300) with corresponding image charges $Q_s'$ and $Q_t'$.

In order to analyze the total Coulomb force influencing the test charge, the concept of the escape energy $E_{escape}$ is introduced. It describes the energy input needed to move $Q_t$ into infinity, and can be related to the difference in potential energy $\Delta E_{pot}$ of the test charge at position $P_a$ and the reference point at $P_\infty$ at infinity:

$$E_{escape} = -\Delta E_{pot} \tag{3}$$

The escape energy $E_{escape}$ is calculated by integrating the effort to counteract the total Coulomb force $\vec{F}$ affecting the test charge as shown in Eq. 4.

$$E_{escape} = -\int_{P_a}^{P_\infty} \vec{F} \cdot d\vec{s} = -Q_t \int_{P_a}^{P_\infty} \vec{E} \cdot d\vec{s}$$
$$= Q_t[\varphi_{total}(P_\infty) - \varphi_{total}(P_a)] = -Q_t \cdot \varphi_{total}(P_a) \tag{4}$$

Here $\vec{E}$ describes the electric field. The electrical potential at $P_\infty$ is considered to be zero. It becomes evident that the escape energy $E_{escape}$ of the test charge can be calculated by the multiplication of the test charge $Q_t$ with the total electrical potential $\varphi_{total}$ at position $P_a$. The potential $\varphi_{total}$ is analytically calculated by a superposition of the potential of the stationary charge $Q_s$ as well as the image charges $Q_s'$ and $Q_t'$ ($\varphi_{total} = \varphi_s + \varphi_{s'} + \varphi_{t'}$). Here the specific potentials $\varphi_i$ are calculated as:

$$\varphi_i = \frac{Q_i}{4\pi\varepsilon_0\varepsilon_r r} \tag{5}$$

Please note that the presented formalism for the total potential $\varphi_{total}$ bears a minor subtlety owing to the simultaneous movement of the image charge $Q_t'$ during the "escape" of the test charge $Q_t$. Even though a direct contribution from $Q_t$ will alter the total potential $\varphi_{total}$ according to its changing position, and therefore a changing position of $Q_t'$, it is not relevant whether the image charge $Q_t'$ is doing work or not. Here, we are focusing on the escape process only, leaving aside the changing energy input (due to polarization) into the high-$k$ material. The result of this escape energy evaluation is presented in the following for the cases A and B.

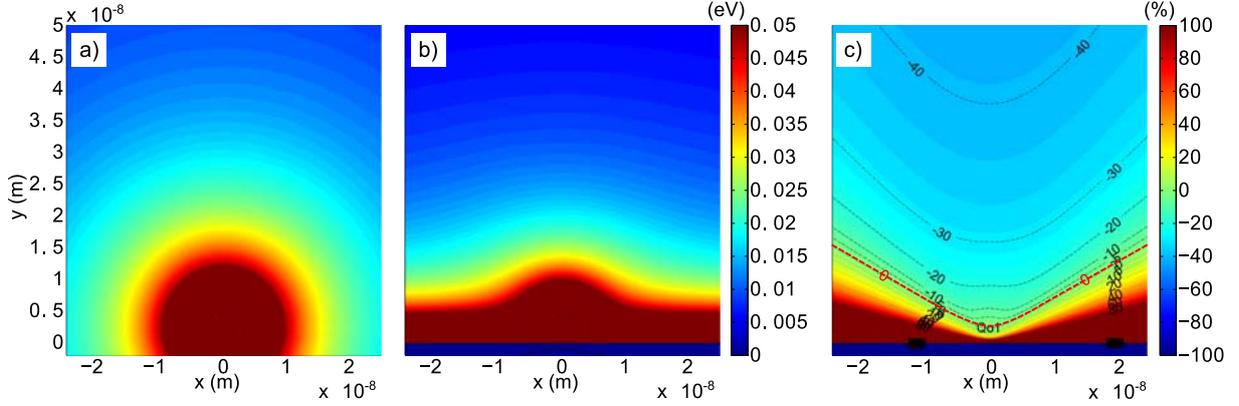

**Figure 2** (a) Escape energy for $Q_t$ at a low-$k$ / low-$k$ ($\varepsilon_r$ = 3 / 3) interface, (b) Escape energy for $Q_t$ at a low-$k$ / high-$k$ ($\varepsilon_r$ = 3 / 300) interface, (c) relative change in the escape energy between the cases a) and b).

Illustrated in Fig. 2a is the escape energy of the test charge for case A. The radial escape energy dependence subject to the distance between test and stationary charge is the result of the homogeneous material permittivity of 3 at the low-$k$ / low-$k$ interface. The escape energy distribution changes significantly for case B, where the low-$k$ / high-$k$ interface ($\varepsilon_r$ = 3 / 300) is considered. Here, the escape energy distribution is to a large degree dependent on the distance between the test charge and the stationary charge as well as the interface (Fig. 2b). The influence of the interface is a consequence of the interaction between the test charge $Q_t$ and the image charges $Q_t$' and $Q_s$'.

The relativ change of the escape energy between case A and B is quantified by Fig. 2c. The graph demonstrates that the benefit of a reduced escape energy due to the introduction of a high-$k$ interface increases, with an increase in the test charge / interface distance. At a distance of 20 nm to the stationary charge and perpendicular to the high-$k$ interface a reduction of the escape energy by 30 % is found, while at a distance of 45 nm the escape energy is reduced by over 40 %, for the used example. For case B the reduced interaction between the stationary charge and the test charge can be ascribed to a screening of the stationary charge $Q_s$ by its image charge $Q_s$'. Here this screening results into an effective $Q_s$ of:

$$Q_{s,effective} = Q_s + Q_s' = \left(1 + \frac{\varepsilon_{r,low-k} - \varepsilon_{r,high-k}}{\varepsilon_{r,low-k} + \varepsilon_{r,high-k}}\right) \cdot Q_s \qquad (6)$$
$$\approx 0.02 \cdot Q_s$$

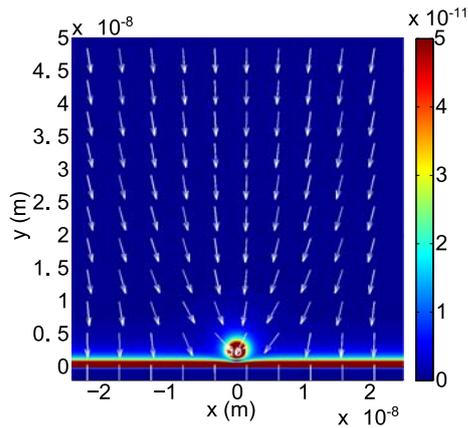

**Figure 3** Force affecting $Q_t$ at a low-*k* / high-*k* interface ($\varepsilon_r$ = 3 / 300). White arrows indicate the direction of the force.

While this screening effect may be beneficial for test charges at a distance to the high-*k* interface beyond the red dashed 0 % relative escape energy change line in Fig. 2c, an enhancement in the escape energy is found for test charges close to the high-*k* interface due to image charge interaction. By considering Fig. 3 however, which illustrates the direction of the Coulomb force for case B, it becomes clear that due to the enhanced interaciton of the test charge with the high-*k* interface, the Coulomb force has a major component perpendicular to the interface. In consequence charge carrier transport in parallel to the high-*k* interface does not require additional energy and will therefore most likely not limit the concept with respect to its organic photovoltaic applicability.

**4 Device Concepts and First Devices** In order to make use of the introduced concept for organic solar cell applications, structures as shown in Fig. 4 are suggested. Here, the active layer is made up of a hybrid structure consisting of an organic acceptor / donator system with integrated inorganic electrically insulating, nanostructured high-*k* materials.

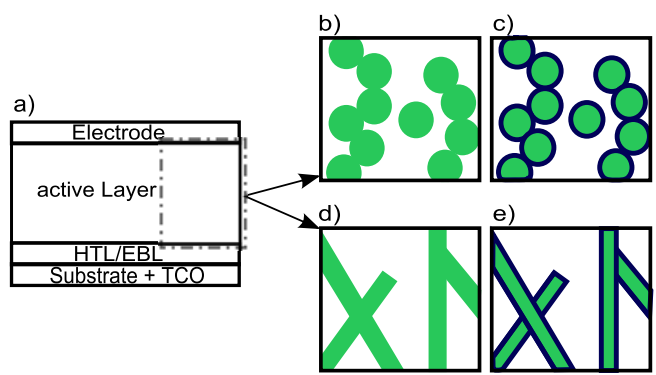

**Figure 4** Organic bulk-heterojunction photovoltaic cell (a) typical device structure, (b) with integrated high-*k* percolating nanoparticles; (c) case (b) modified with an organic coating; (d) with integrated nanowires and (e) case (d) modified with organic coating.

Amongst other implementation possibilities the concept can be realized by the integration of either high-*k* nanoparticles or nanowires. As long as percolation along the particles or nanowires between the respective electrodes of the organic solar cell is possible, the structure should not be limited by the discussed charge carrier / high-*k* interface interaction. To insure a good dispersability of the nanostructures in an organic solvent alongside the acceptor / donor material and to positively influence the thin film morphology, a functionalization might become nessecary. However, due to the demonstrated long reaching influence of the charge carrier screening, a functionalization should not limit the concept. The structures as described so far may however result in recombination losses as illustrated in Fig. 5, where the high-*k* nanostructure connects acceptor and donor phases and as such may lead to the trapping of complementary charges. Encasing the nanostructures with either an organic acceptor or donor material prior to dispersion, as suggested by Fig. 4c) and Fig. 4e) may therefore further improve the discussed approach.

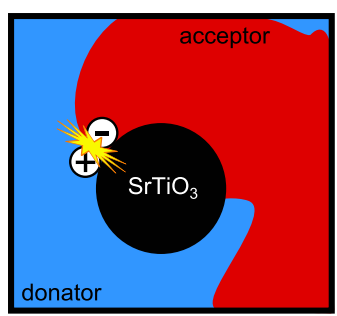

**Figure 5** Limit to the concept by recombination when high-*k* nanostructures interconnect acceptor / donor phases.

To test the feasibility of the discussed concept for an organic photovoltaic application, first P3HT:PCBM devices containing high-*k* $SrTiO_3$ nanoparticles were implemented according to approach A as described in the experimental section. The nanoparticle distribution in the organic active layer is illustrated by the thin film top view scanning electron microscopy (SEM) image shown in Fig. 6. The shown $SrTiO_3$ agglomerates build up percolation paths through the active layer, which enable charge carrier transport to the respective electrodes. Over the whole sample area, the agglomerates are evenly distributed (not shown). Further thin film morphology improvements are ongoing research.

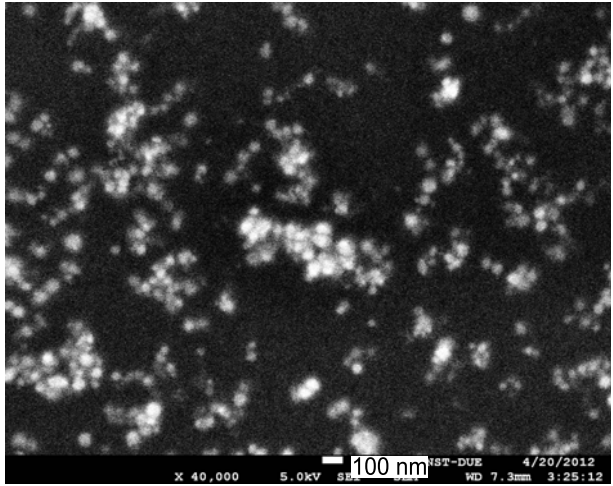

**Figure 6** SEM image of the P3HT:PCBM thin film with integrated $SrTiO_3$ nanoparticles.

Illustrated in Fig. 7a) is the current voltage characteristic of a P3HT:PCBM:$SrTiO_3$ device in comparison to a
P3HT:PCBM reference. The devices were measured in the dark as well as under AM1.5 standard illumination. In comparison to the reference device, the short circuit current for the high-$k$ cell has been reduced by ~3.6 % from $J_{sc}$ = 7.9 mA cm$^{-2}$ down to $J_{sc}$ ~ 7.63 mA cm$^{-2}$. The fill factors (*FF*) on the other hand, as well as the open circuit voltages ($V_{oc}$) have been improved by ~ 10 %. The *FF* shows an improvement from 0.49 (reference) up to 0.54 (high-$k$), while the open circuit voltage has been increased from $V_{oc}$ = 0.52 V (reference) up to $V_{oc}$ = 0.57 V (high-$k$). In terms of power conversion efficiency, this translates into an efficiency improvement for the high-$k$ solar cell of ~17 % from $\eta$ = 2 % absolute for the reference cell up to $\eta$ = 2.34 % absolute for the high-$k$ cell. It should be noted that the obtained absolute efficiencies of ~ 2 % for both reference and high-$k$ devices are lower than one would expect for a standard P3HT:PCBM system, where efficiencies in the order of ~ 3.5 % are a good average [16, 17]. This deviation is ascribed to the extensive stirring times of ~ 24 h at $T$ = 65-70 °C during the preparation of the P3HT:PCBM solution. The extensive stirring time was required to disperse the $SrTiO_3$ nanoparticles and as such may lead to a degradation of the organic semiconductor or morphology effects which possibly degrade the short circuit current. However, since both the reference cells as well as the high-$k$ cell were realized with solutions / dispersions exposed to identical stirring times, this preliminary experimental circumstance still allows for an interpretation of the influence of the high-$k$ nanoparticles on the organic solar cell performance.

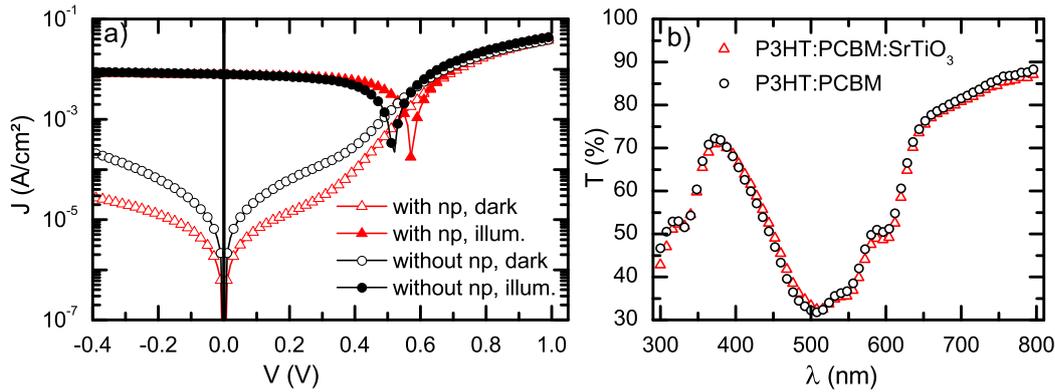

**Figure 7** (a) $I(V)$-characteristics of P3HT:PCBM:SrTiO$_3$ solar cells and of P3HT:PCBM reference devices, (b) Transmission spectrum of P3HT:PCBM:SrTiO$_3$ and P3HT:PCBM thin films.

By considering Fig. 7b), which illustrates the result of a transmission measurement experiment on P3HT:PCBM thin films with and without SrTiO$_3$ nanoparticles, we find that the thin film transmission does not change due to the nanoparticle integration. Conducted reflection measurements (not shown) yield the same result. This allows us to conclude that the obtained power conversion efficiency improvement for the organic solar cell with integrated nanoparticles is not the result of light-scattering effects and as such an enhanced optical path length within the cell. Further, by observing that $J_{sc}$ is reduced and $V_{oc}$ is enhanced upon nanoparticle integration, we can exclude beneficial morphology effects as a cause for the enhanced efficiency. We conclude the enhanced power conversion efficiency of the solar cell with integrated SrTiO$_3$ particles to be the result of an effectively enhanced device permittivty. The development of $J_{sc}$ and $V_{oc}$ upon nanoparticle integration supports this statement. For P3HT:PCBM organic solar cells, with a very efficient exciton separation at the acceptor / donor heterointerface, a low geminate recombination probability was demonstrated by Mingebach et.al. [16]. Therefore no increase in $J_{sc}$ for an enhanced effective system permittivity is to be expected. An increase of $V_{oc}$ for an enhanced system permittivity on the other hand is expected. This is ascribed to a reduction of the non-geminate recombination probability during charge carrier transport to the respective electrodes, as the result of a reduced Coulomb interaction. This can be understood by describing the non-geminate recombination during charge transport as a low shunt resistance within the solar cell. In consequence, a reduction in the non-geminate recombination probability will lead to an increase of the shunt resistance and therefore into an increased solar cell open circuit voltage.

**5 Conclusions** In conclusion we introduced a new concept to enhance the effective permittivity of organic semiconductor thin films, by the integration of electrically insulating, nanostructured high-*k* materials. This concept is intended for an application in organic photovoltaic cells. Using an analytical model we substantiated that an enhancement of the effective organic thin film permittivity will reduce the Coulomb interaction in the organic layer, and as such will improve charge carrier transport properties. This was experimentally verified by realizing first P3HT:PCBM:SrTiO$_3$ nanoparticle test devices, which demonstrate an enhanced power conversion efficiency of ~ 17 % in comparison to a P3HT:PCBM reference device. This improvement is ascribed to a reduction in non-geminate recombination, which is in line with the influence of an effectively enhanced organic layer permittivity.

**Acknowledgements** The authors acknowledge funding within the framework of a North Rhine-Westphalia scholarship for "rollable solar cells". Further, the group of Prof. G. Bacher, located at the University of Duisburg-Essen, is acknowledged for making the transmission and reflection measurements possible.